\documentclass[a4paper,preprint,showpacs,amssymb,superscriptaddress]{revtex4}
\usepackage{graphicx}

\begin{document}
\draft
\title{A complexity measure for symbolic sequences and applications to DNA}
\author{Ana P. Majtey}
\address{Facultad de Matem\'atica, Astronom\'\i a y F\'\i sica \\
Universidad Nacional de C\'ordoba \\ Ciudad Universitaria, 5000
C\'ordoba, Argentina \\ CONICET}
\author{Ram\'on Rom\'an-Rold\'an}
\address{Departamento de F\'isica Aplicada, Universidad de Granada, Granada,
Spain}
\author{Pedro W. Lamberti}
\address{Facultad de Matem\'atica, Astronom\'\i a y F\'\i sica \\
Universidad Nacional de C\'ordoba \\ Ciudad Universitaria, 5000
C\'ordoba, Argentina \\ CONICET}
\begin{abstract}
We introduce a complexity measure for symbolic sequences. Starting
from a segmentation procedure of the sequence, we define its
complexity as the entropy of the distribution of lengths of the
domains of relatively uniform composition in which the sequence is
decomposed. We show that this quantity verifies the properties
usually required for a ``good'' complexity measure. In particular
it satisfies the one hump property, is super-additive and has the
important property of being dependent of the level of detail in
which the sequence is analyzed. Finally we apply it to the
evaluation of the complexity profile of some genetic sequences.
\end{abstract}


\date{\today}
\maketitle

\section{Introduction}

\noindent In the last few years the term {\it complexity} has
become frequent in scientific literature \cite{Bennett, Kauffman,
Sole}. This has conveyed the introduction of diverse complexity
measures in different areas of science. Kolgomorov's algorythmic
complexity \cite{Kolmogorov}, Lempel \& Ziv's measure
\cite{Lempel}, Bennet's thermodynamic depth
\cite{Bennett},\cite{Lloyd}, physical complexity \cite{Adami} or
Lopez-Ruiz, Mancini \& Calvet's complexity measure \cite{LMC}, are
some of the examples that have caught most attention. In fact,
this list does not reflect all the proposed complexity measures.

In spite of these efforts, and reflecting such diversity,
consensus is to be reached about a precise definition of the
complexity concept that would allow its quantification. It is
possible that one of  the main difficulties to reach that
consensus is the lack of a language that is common to all the
different areas of science in which the concept is meant to be
introduced. As an example, the notion of information and its
quantifier, the entropy, is usually present in measures proposed
to evaluate the complexity of a system or of a process. At the
same time, entropy, in physics is a measure of the disorder of the
system, which grows as the disorder grows. However, intuitively, a
complex system may simultaneously involve order as well as
disorder. Two extreme cases are to be considered when, in physics,
a complexity measure is searched. Firstly, a perfect crystal (a
completely ordered system) and on the other hand the ideal gas (a
completely disordered system). Clearly both systems have no
complexity (or an extremely low complexity). In general, a
properly defined complexity measure should reach its maximum at
some intermediate level between the order of the completely
regular and the disorder of the absolutely random. This desirable
characteristic for all complexity measures is known as the {\it
one hump} property.

Very often, a complex system is described as one formed by many
non-lineal elements that interact with each other \cite{chialvo}.
These interactions give the system the capacity to auto-organize
\cite{Bak}. Given the fact that complexity comes from the
interactions of the single units, these interactions must be taken
into account when defining a measure that quantifies the
complexity of a system. When the different parts of a system,
e.g., the molecules of an ideal gas in equilibrium, do not
interact, their behavior can be understood as the sum of its
separated components. But, when interdependencies occur, this is
not valid anymore and to quantify the complexity we need a measure
that takes those bonds into consideration \cite{Sole}.

An adequate complexity measure should be super-additive, meaning
that the two systems' juxtaposition gives as a result a system in
which complexity equals or exceeds the addition of the considered
systems. This means that the (extensive) complexity of the {\it
whole} is equal or larger than the sum of the (extensive)
complexities of the \textit{parts}. Here we are devoted to
investigate a complexity measure for symbolic sequences. In this
case, the super-additive property reads as follows: if
$C_{{\cal{S}}_1}$ and $C_{{\cal{S}}_2}$ denote the complexities of
two symbolic sequences ${\cal{S}}_1$ and ${\cal{S}}_2$,  with
corresponding lengths $L_1$ and $L_2$, then
\begin{equation}
(L_1 + L_2) \; C_{{\cal{S}}_1 {\cal{S}}_2}\geq L_1 \;
C_{{\cal{S}}_1} + L_2 \; C_{{\cal{S}}_2} \label{supad}
\end{equation}
where $C_{{\cal{S}}_1  {\cal{S}}_2}$ denotes the complexity of the
juxtaposition of ${\cal{S}}_1$ and ${\cal{S}}_2$.

The complexity measure we introduce in the present work takes into
account the lengths of the segments of relatively  uniform content
in which a symbolic sequence is divided. To establish the
segmentation we must look for compositionally homogeneous
segments. Then, two extreme cases may occur after the segmentation
process:

\begin{itemize}
\item  all the resulting segments have the same length (periodic
sequence),
\item  the sequence has not been segmented (random
sequence).
\end{itemize}
These two cases correspond with the perfect crystal and the ideal
gas mentioned earlier, and as we will see, they have a null
complexity, according to our definition. Now the next step is to
characterize what we will take as the \textit{most} complex
sequence, that is, we must fix a third point over the complexity
plot. In order to do that, we go along the following line of
reasoning: when the probability, of measuring a particular value
of a certain quantity, varies inversely as a power of that value,
it is said that the quantity follows a power law. The importance
of the distributions following a power law in physics and related
areas has been pointed out by the ubiquity of such laws in a wide
range of phenomena. This type of laws rules as much the frequency
of the use of words in any human language as the number of moon
craters of  a particular size \cite{Newman}. In general it is
accepted that a power law dependence is an indication of
hierarchical organization. More interestingly, this kind of
behavior also appears in brain dynamics studies. In fact, it is
known that the brain constantly makes complex functional nets
corresponding to the traffic between regions. In this case it is
found that the probability for $k$ regions to be temporarily
correlated with a given region satisfies a rule  $k^{- \mu}$ where
$\mu \approx 2$ \cite{chialvo2}. To us, this example proves to be
highly significant because brain dynamics is a milestone case of
auto-organization and undoubtedly of what we can consider as a
complex system. At its time, auto-organization is seen as the
modelling mechanism to a great amount of systems in Nature.

According to these precedents, we consider reasonable to take as a
high complexity sequence, one that has a lengths distribution of
patches of relatively uniform composition following a power law,
i.e. the probability $P(l)$ of finding a patch of relatively
homogeneous composition with length $l$, is given by:
\begin{equation}
P(l) \sim \frac{1}{l^{\mu}}. \label{pow1}
\end{equation}
We suppose further that the most complex sequence is the one in
which the interdependence between subsegments is maximum. To
quantify that interdependence, we use the autocorrelation
function, $C(l)$ \cite{Stanley}. Interdependence is maximum when
the autocorrelation function is flat. There exists an interesting
relationship between the exponent $\mu$ in (\ref{pow1}), and the
behavior of the autocorrelation function \cite{Stanley}. In fact,
for a length distribution law given by (\ref{pow1}) it has been
shown that the standard deviation in the symbol content of the
sequence, $F(l)$, has a behavior of the form
$$ F(l) \sim l^{\alpha}
$$
and the autocorrelation function follows a power law
$$
C(l) \sim \frac{1}{l^{\gamma}}
$$
with $\gamma=2-2\alpha$. For an exponent $\mu\leq 2$ corresponds
an exponent $\alpha =1$ and therefore $\gamma=0$, that is, a flat
autocorrelation function \cite{Stanley}. Thus, for extremely long
sequences a flat autocorrelation is associated to a segments
lengths distribution that complies with a power law in which
$\mu\leq 2$. It should be emphasized that every exponent $\mu\leq
2$ leads to a flat autocorrelation function. However the exponent
$\mu=1$ corresponds to a statistically self similar distribution
of patches along the sequence \cite{GCRO}. These facts suggest us
to take as the most complex sequence the one with a lengths
distribution of patches of relatively uniform composition is given
by the law (\ref{pow1}) with $\mu=1$.

This work is organized as follows: In Section II we describe the
sequence segmentation method implemented; in Section III we
introduce a complexity measure and study its basic properties; in
Section IV we apply the introduced measure to real genomic
sequences; finally we present some conclusions.

\section{Segmentation method}

\noindent In this section we describe the segmentation algorithm
applied to the study of the sequence structure. The method is
based on the Jensen-Shannon entropic divergence (JSD) and it was
successfully applied to the study of DNA sequences \cite{RRR}. DNA
sequences are formed by patches or domains of different nucleotide
composition; given the huge spatial heterogeneity of most genomes,
the identification of compositional patches or domains in a
sequence is a critical step in understanding large-scale genome
structure \cite{SCC}.

The JSD is a measure of distance between probability
distributions. Although it was initially  defined as a distance
between two probability distributions, Lin has proposed a
generalization to several probability distributions \cite{Lin}.
Let $P^{(k)}=\{{p_i^{(k)}, i=1..N\}, k=1..M}$, a set of $M$
probability distributions $(\sum_i p_i^{(k)}=1, \;k=1..M)$, for a
discrete variable $X$ with $N$ possible values $X_i$; $p_i^{(k)}$
denotes the probability of occurrence of the value $X_i$ according
to the distribution $P^{(k)}$. The JSD for these probabilities
distributions is defined by:
\begin{equation}
JS[P^{(1)},.., P^{(M)}]=H[\sum_k \pi^{(k)}P^{(k)}]- \sum_k^M
\pi^{(k)}H[P^{(k)}] \label{JS}
\end{equation}
where $H[P]=-\sum_j p_j \log_2 p_j$ is the Shannon's entropy and
the numbers $\pi^{(k)},\;k=1..M,\;\sum_k \pi^{k}=1$ are weights
properly chosen.

The JSD is non negative, bounded and can be interpreted in the
frame of information theory \cite{Grosse}. Incidentally we mention
that the JSD has been proposed as a complexity measure for genomic
sequences \cite{SCC}.

In the context of symbolic sequences analysis, the probabilities
$p_i$ are approximated by the frequency of occurrence of each
symbol throughout the sequence. For a DNA sequence, the symbols
are the nucleotides $\{A;C;T;G\}$. If we want to compare the
compositional content of two symbolic sequences, let us say
${\cal{S}}_1$ and ${\cal{S}}_2$, of lengths $L_1$ and $L_2$, we
can use the expression (\ref{JS}), where the weights are taken
equal to $\pi^{(k)}=L_{k}/L$, $k=1,2$, with $L=L_1+L_2$. In this
case the probability  distributions $P^{(1)}$ and $P^{(2)}$ are
approximated by the frequency of occurrence of the different
symbols throughout each sequence.

The segmentation procedure allows to decompose the sequence into
domains or subsequences with a different base composition in
comparison to the two adjacent subsequences, at a given level of
statistical significance or threshold, $D_u$. This threshold is
associated with the level of details in which the sequence is
analyzed \cite{Grosse}.

In order to make this paper self-contained we will describe the
basic steps in the segmentation procedure. For a more detailed
description we refer the reader to reference \cite{RRR}. Let us
suppose that we define a moving cursor along the complete
sequence. For each position of the cursor, it results two
subsequences, one to the left and other to the right of the
cursor. For each subsequence we can evaluate the occurrence
frequency of each symbol and then calculate the JSD for each
position of the cursor. The position that corresponds to a maximum
of the JSD above the threshold elected, $D_u$, is taken as a cut
point. Clearly these points corresponds to the maximum of the
discrepancy between the compositional content of each
subsequences. The procedure is repeated for each resulting
subsequence until the JSD be greater than the threshold value.

When segmenting symbolic sequences with simple domain structures,
homogeneous domains can be consistently found (if purely random
fluctuations are excluded). However, when the method is applied to
long-range correlated sequences, such homogeneity vanishes: by
relaxing the threshold value, we find new domains within other
domains, previously taken as homogeneous under a higher threshold
value. This domains-within-domains phenomenon points to complex
compositional heterogeneity in DNA sequences, which is consistent
with the hierarchical nature of biological complexity \cite{SCC}.
We will back to this point at the end of the present work.

\section{Definition of the complexity}

Let us consider a symbolic sequence $\cal{S}$ of length $L$ (i.e.,
$L$ is the number of symbols in the sequence). Let us assume that
by segmenting the sequence according to procedure described in the
preceding section, we can decompose the sequence in $N_s$ patches
or domains of different compositional content (up to a
significance level $D_u$) \cite{Grosse}. Let us denote by $l_i,
i=1...N_s$, the lengths of each one of these segments. Obviously
\begin{equation}
\sum_{i=1}^{N_s}l_i = L
\end{equation}
In general these lengths are not all different. Let us denote by
$\Omega$ the subset of lengths $l_i$ such that $l_i\neq l_j$ if $i
\neq j$:
\[
\Omega = \{(l_{\alpha_1},...,l_{\alpha_\kappa}), l_{\alpha_i} \neq
l_{\alpha_j} \;\; if \;\; i \neq j, \kappa\leq N_s\}
\]
Let $N_{\alpha_i}$ be the number of segments of length
$l_{\alpha_i}$. Then $\sum_{i=1}^\kappa N_{\alpha_i} = N_s$. Let
us consider now an arbitrary partition ${\cal{A}} =
\{A_j\}_{j=1}^\nu$, of the interval $[1,L]$ with $\nu - 1$ (the
number of subintervals), in principle, arbitrary:
\begin{equation}
1 = A_1 < A_2 < ...< A_{\nu-1} < A_\nu =L
\end{equation}
We name the quantity $\Delta_j=A_j - A_{j-1} \;\; j=2,...,\nu$ as
the amplitude of the corresponding subinterval.

Let us denote by $\tilde{N}_j$ the number of patches in the
segmented sequence with length belonging to the interval
$[A_{j-1},A_j)$. The condition $\sum_{j=2}^{\nu} \tilde{N}_j =
N_s$ is satisfied. Finally let us denote by $f_j$ the occurrence
frequency of segments whose length belongs to the interval
$[A_{j-1},A_j)$ (with the convention that the interval
corresponding to $j=\nu$ includes the extreme value $L$):

\begin{equation}
f_j =\frac{\tilde{N}_j}{N_s}; \;\;\; \sum_{j=2}^{\nu} f_j =1
\label{freqqq}
\end{equation}

From the knowledge of the frequencies $F = \{f_j\}$ we can
evaluate the Shannon's entropy
\begin{equation}
H_{{\cal{S}}}(F; {\cal{A}}, D_u)  \equiv H[F] = -\sum_{j=2}^\nu
f_j \log_2 f_j \label{fund}
\end{equation}
Clearly this quantity depends on the partition $\cal{A}$, and on
the significance level $D_u$ at what the segmentation was done,
that is, it depends on the level of detail at what the sequence
was analyzed. Therefore we have included explicitly the partition
$\cal{A}$ and the significance value $D_u$ as arguments in
$H_{{\cal{S}}}$.

There are two cases in which the entropy (\ref{fund}) does not
depend on the particular partition chosen:
\begin{enumerate}
\item a \textsl{idealized} periodic sequence and
\item a \textsl{idealized} random sequence.
\end{enumerate}
Here what is meant by \textsl{idealized} is that the respective
character is detected to every significant level of detail of the
analysis. In the first case, there exists only one value (the
period) for the length of the segments. Therefore $f_J=1$ for some
value $2\leq J \leq \nu$ and $f_j=0$ for all other $j$. Thus, for
a periodic sequence $H_{\cal{S}}=0$ for any partition of the
interval $[1,L]$. Analogously, due to the fact that a random
sequence is not segmented at any significant level of detail (by
the proper meaning of significant), only one of the $f_j$ is
different of zero: $f_\nu=1$. Thus we also have $H_{\cal{S}}=0$ is
this case. These two extreme cases are the corresponding ones with
the crystal and the isolated ideal gas, in the physical context.
In that sense, $H_{{\cal{S}}}(F; {\cal{A}}, D_u)$ is a good
candidate as a complexity measure. It should be emphasized that
$H_{\cal{S}}$ has information about the segmentation of the
sequence. The fact that $H_{\cal{S}}$ vanishes for a periodic and
a random sequence, suggests to investigate it as a measure of
complexity. However, it should be also indicated that, in order to
be a true characteristic of the sequence under study, a complexity
measure must be independent of any  arbitrary parameter. For it, a
particular partition is adopted by refining the complexity measure

Now we proceed to characterize, in a formal way, what we will take
as the most complex sequence. Let us assume that after the
segmentation procedure, at a given level of detail, the sequence
${\cal{S}}$ is decomposed in $N_s$ segments of uniform
compositional content, and let us suppose that we are able to
identify a power law for the distribution of the segments length:
\begin{equation}
N_l = \frac{N_s}{Z(\mu,\lambda^*)} l^{-\mu} \label{power}
\end{equation}
where $Z(\mu,\lambda^*) = \sum_{l=1}^{\lambda^*} l^{-\mu}$,
$\lambda^*$ is a cutoff length and $\mu \geq 1$. As we indicated
in the introduction and for the reasons there expressed we chose
$\mu =1$. The cutoff $\lambda^*$ have to do with the finite size
of the sequence ${\cal{S}}$. Its value can be deduced from the
condition
\begin{equation}
N_s \frac{Z(\mu-1,\lambda^*)}{Z(\mu,\lambda^*)}=L
\end{equation}
From the distribution law (\ref{power}), and for a given partition
$\cal{A}$, we can evaluate the frequencies
\begin{equation}
f_j= \frac{1}{N_s} \sum_{l \epsilon [A_j, A_{j+1}-1]} N_l,
\label{freqq}
\end{equation}
and from these one, the entropy (\ref{fund}).

At this point we look for the partition ${\cal{A}}$ that makes the
entropy (\ref{fund}) to reach a maximum value when the frequencies
(\ref{freqq}) are replaced. Due to a fundamental property of the
entropy, the maximum value of $H_{{\cal{S}}}(F; {\cal{A}}, D_u)$
is reached for a partition $\cal{A}$ such that all the frequencies
$f_j$ are equal for all $j$, that is, the number of segments
belonging to the interval $[A_{j-1}, A_j)$ is the same for all
$j$. Due to the cutoff, there exists a value $j^*$ such that
$f_j=0$ for $j>j^*$. Hence, the maximum of the entropy corresponds
to the biggest $j^*$ consistent with the uniformity condition for
the $f_j$. The entropy $H_{{\cal{S}}}(F; {\cal{A}}, D_u)$ will be,
in this case, $\log_2 j^*$.

To satisfy the above two conditions, that is, the uniformity of
$f_j$ for $j\leq j^*$ and the biggest value for $j^*$, we must
find a partition ${\cal{A}}$ of the interval $[1,L]$ such that the
number of segments in each interval is constant and equal to one.
These requirements can be expressed as a set of equations to be
satisfied by the extremes of each one of the intervals of the
partition ${\cal{A}}$:
\begin{eqnarray}
1+\frac{1}{2^{\mu}}+\ldots+\frac{1}{(A_2-1)^{\mu}} & = &
\frac{1}{A_2^{\mu}}+\ldots+\frac{1}{(A_3-1)^{\mu}} \nonumber \\
\nonumber \frac{1}{A_3^{\mu}}+\ldots + \frac{1}{(A_4-1)^{\mu}}& =
& \frac{1}{A_4^{\mu}}+\ldots +
\frac{1}{(A_5-1)^{\mu}} \nonumber \\
\vdots  \label{partition} \\
\frac{1}{A_{j^*-2}^{\mu}}+\ldots + \frac{1}{(A_{j^*-1}-1)^{\mu}} &
= & \frac{1}{A_{j^*-1}^{\mu}}+\ldots + \frac{1}{(\lambda^*)^{\mu}}
\nonumber
\end{eqnarray}
with $\mu=1$.

As we are looking for the maximum $j^*$ it is obvious from the
previous set of equations that we must take $A_2=2$. The rest of
the amplitudes $\Delta_j= A_j - A_{j-1}$  can be obtained from the
set of equations (\ref{partition}).

Now we are in position to introduce our complexity measure for an
arbitrary symbolic sequence ${\cal{S}}$ of length $L$. We define
it as:
\begin{equation}
C_{\cal{S}}=H[F_L], \label{basic}
\end{equation}
where $H[F_L]$ is the entropy of the distribution of lengths of
the domains in which the sequence has been decomposed, evaluated
according to the partition of the interval $[1,L]$ given by the
relations (\ref{partition}) with $\mu=1$.

The evaluation of complexity (\ref{basic}) for an arbitrary
sequence ${\cal{S}}$ of length $L$ requires:
\begin{enumerate}
\item To calculate the partition ${\cal{A}}$ corresponding to the
length $L$ according to (\ref{partition}) for $\mu=1$;
\item by using the segmentation procedure described in section II,
at certain significance value $D_u$, evaluate the set of length
$\Omega$ and from it the frequencies $f_j$ given by (\ref{freqqq})
for the partition ${\cal{A}}$;
\item finally, evaluate the entropy $H_{{\cal{S}}}$ given by
(\ref{fund}).
\end{enumerate}

Incidentally it is worth to mention that for a greater value of
$\mu$ compatible with the flat autocorrelation condition $(\mu\leq
2)$, the entropy $H[F_L]$ evaluated following the previously
described steps, takes values extremely slow. Therefore, besides
the conceptual motives that led to the election of $\mu =1$, there
are practical ones as well.

\section{Applications and results}

In this section we apply the proposed measure to the evaluation of
the complexity for some DNA sequences. In all examples the
quaternary alphabet $\{A,T,C,G\}$ is used. These evaluations allow
us, on one side, to study the main properties of the measure, such
as the dependence with the level of detail in the analysis of the
sequence and the super-additivity property; on the other we can
investigate our measure as an adequate tool for unravelling
certain structural features within the DNA, for instance, the
content of introns and exons, and its relation with evolutionary
aspect of the genome.

As it was already claimed, an appropriate complexity measure
should take into account the level of detail at what the system
under study is analyzed \cite{WLi}. To check this dependence we
apply the measure (\ref{basic}) to real DNA sequences with
different correlation structure and to a computer generated random
sequence. Figure 1 shows the complexity $C_{\cal{S}}$ as a
function of the threshold level, $D_u$, for the genomic sequences
HUMTCRADCV, the ECO110k and the random one (this kind of plots are
known as complexity profile). The first one is a human DNA
sequence with long range correlations \cite{Peng}. The second one
is an uncorrelated bacterial sequence. A first remarkable aspect
of $C_{\cal{S}}$ is that there exists a range for the significance
value $D_u$, $20\leq D_u \leq 50$, for which it gets the null
value when evaluated for the random sequence. This random sequence
has been built with identical composition that those of the
ECO110k. For $D_u$ belonging to this interval, the values of the
complexity for the human sequence are greater than those for the
bacterial one. This fact is consistent with taking as range of
interest for the threshold the interval previously indicated. One
noticeable characteristic of the complexity profiles for the
natural sequences, is that, unlike those obtained for the
complexity measure introduced in \cite{SCC}, do not go to zero as
the threshold $D_u$ increases.

Another investigated aspect of $C_{\cal{S}}$ has to do with the
super-additivity property, eq. (\ref{supad}). In figure 2 we show
the complexity profiles for the complete DNA sequences ECO110k and
the human beta-globulin HUMHBB, and the weighted sum of the
complexity profiles for two arbitrary subsequences of these two
sequences. Clearly the equation (\ref{supad}) is verified. It is
obvious from the definition of $C_{\cal{S}}$ that the complexity
of any self concatenation of an arbitrary sequence is equal to
complexity of the original sequence whenever the fusion point
coincides with a cut point resulting from the segmentation
procedure. If this is not the case, the resulting value for the
complexity of the concatenated sequence might be, for very long
sequences, slightly different to the complexity of the original
sequence.

It is known that only a small portion of the genome of higher
organisms encodes information for amino acid sequences of proteins
\cite{Lewin}. The role of introns (continuous noncoding regions in
DNA) and intergenomic sequences (noncoding DNA fragments
intertwined between coding regions) remain still unknown. The
study of the statistical properties of the noncoding regions has
shown the existence of long range correlations which indicate the
presence of an underlying structural order in the intron  and
intergenomic segments. This structural order is made apparent in
the complexity profiles shown in figure 3, where we have plotted
the complexity values for the coding and noncoding regions of the
human chromosome 22.

Genomic sequences are a valuable source of information about the
evolutionary history of species \cite{huynen}. In particular it
has been possible to relate some statistical characteristics
observed along genomic sequences to the influences of a variety of
ongoing processes including evolution \cite{Buldy}. In this
context we conclude this work evaluating the complexity
$C_{\cal{S}}$ for homologous DNA sequences of different species;
in particular for the myosin heavy-chain. In general it can be
observed that there exists a concordance between the biological
complexity of the species and the values of $C_{\cal{S}}$. It
should be emphasized that there exists a relationship between the
percentage of introns and the long-range correlations in the
sequence. This fact is clearly manifested by the complexity
$C_{\cal{S}}$ as can be observed in figure 4.

\begin{center}
AKNOWLEDGMENT
\end{center}

APM and PWL are grateful to Secretaria de Ciencia y T\'{e}cnica de
la Universidad Nacional de C\'ordoba, Argentina, for financial
assistance. AM is a fellowship holder of CONICET and PWL is a
member of CONICET. This work was partially supported by Grant
BIO2002-04014-C03-03 from Spanish Government. The authors like to
thank Professors Jos\'{e} Oliver and Domingo Prato for useful
comments.

\newpage
\begin{figure}
\begin{center}
\includegraphics[width=15cm,angle=0]{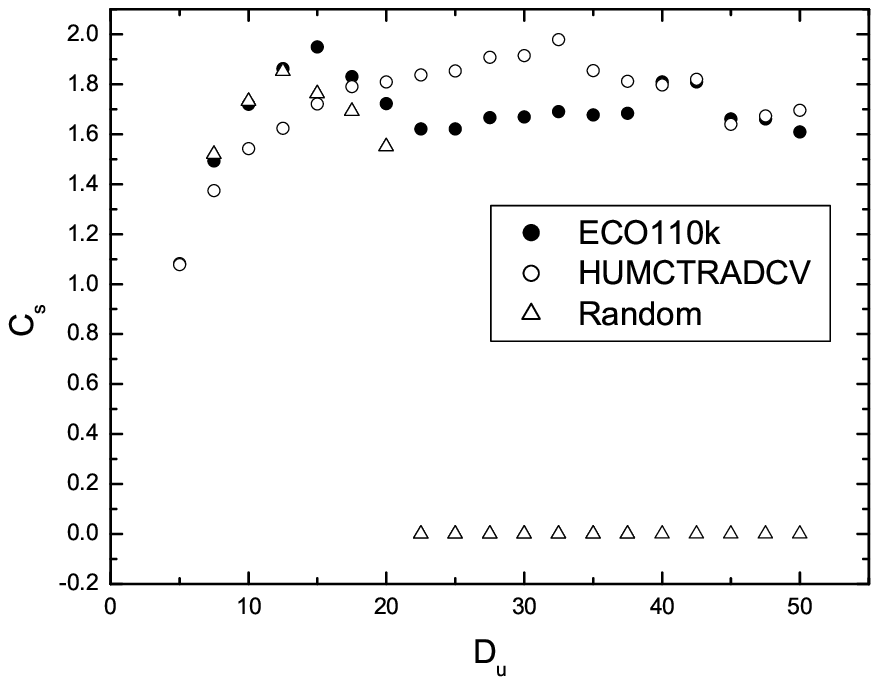}
\end{center}
\caption{Complexity profiles of two natural sequences and a
computer generated random sequence. In this last case, the
sequence has the same compositional content that the ECO110k.}
\end{figure}

\newpage
\begin{figure}
\begin{center}
\includegraphics[width=16 cm,angle=0]{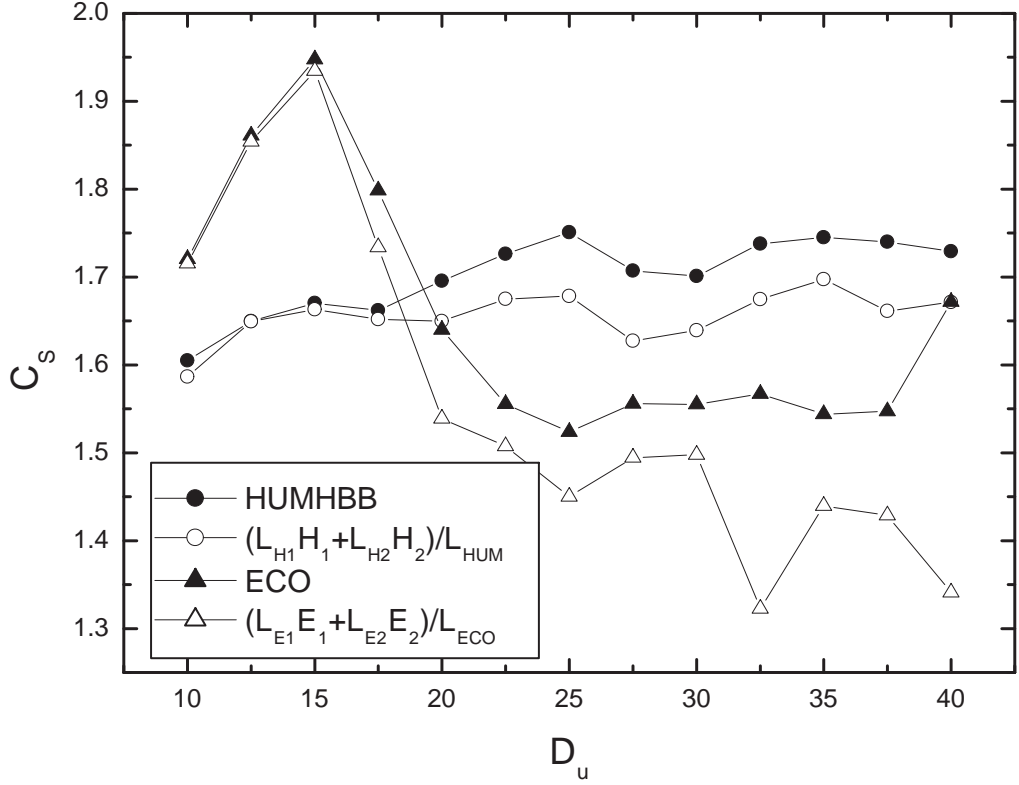}
\end{center}
\caption{Complexity profiles for the sequences ECO110k ($L_{ECO}=
111408$ bp) and HUMHBB ($L_{HUM} = 73308 $ bp). The filled symbols
correspond to the complexity for the whole sequences, and the
empty ones correspond to the (weighted) sum of the complexities
for two arbitrary subsequences of each sequence. The subsequences
were taken in such a way that their juxtaposition were equal to
the complete sequence ($L_{E1} = 57120$ bp and $L_{E2} = 54288$
bp; $L_{H1} = 42720$ bp and $L_{H2} = 30588 $bp).}
\end{figure}

\newpage
\begin{figure}
\begin{center}
\includegraphics[80mm,0mm][25mm,80mm]{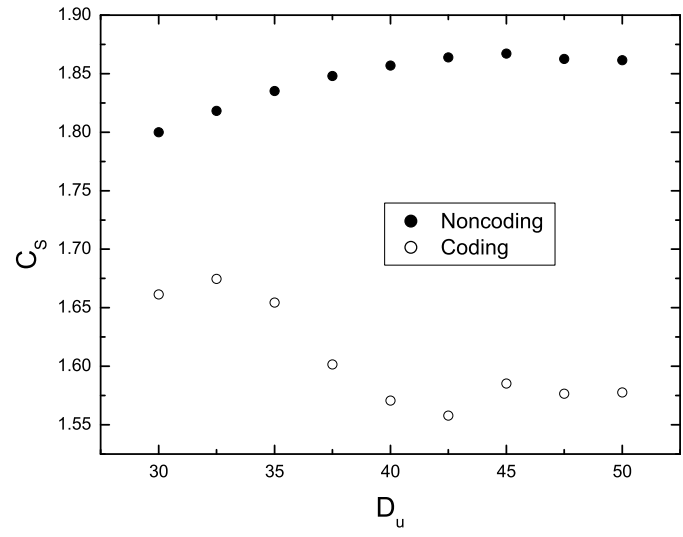}
\end{center}
\caption{Differences in $C_{\cal{S}}$ between coding and noncoding
regions of the sequence corresponding to human chromosome 22.}
\end{figure}

\newpage
\begin{figure}
\begin{center}
\includegraphics[width=13cm,angle=0]{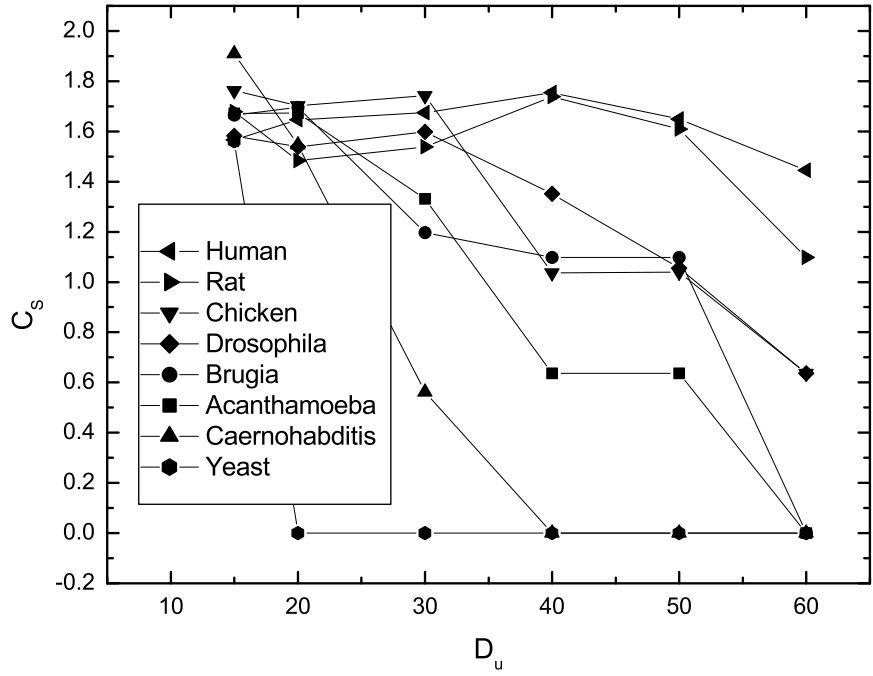}
\end{center}
\caption{Complexity profiles of myosin heavy-chain genes in
different species (total length, percentage of introns): Human
(28438bp, 74\%), Rat (25759bp, 77\%), Chicken (31111bp, 74\%),
Drosophila (22663bp, 66\%), Brugia (11766bp, 32\%), Acathamoeba
(5894bp, 10\%), Caenorhabditis (10780bp, 14\%), Yeast (6108bp,
0\%)}
\end{figure}

\end{document}